\documentclass[conference,a4paper]{IEEEtran}

\usepackage{amssymb}
\usepackage{amsmath}
\usepackage{graphicx}
\usepackage{citesort}

\newcommand{\figref}[1]{Fig.~\ref{#1}}
\newcommand{\figsref}[2]{Figs.~\mbox{\ref{#1}--\ref{#2}}}
\newcommand{\eqsref}[2]{\mbox{(\ref{#1})--(\ref{#2})}}

\newcommand{\sg}{\sigma}
\newcommand{\ld}{\lambda}
\newcommand{\g}{\gamma}
\newcommand{\mhat}{\hat{\mu}}
\newcommand{\shat}{\hat{\sg}}
\newcommand{\ghat}{\hat{\g}}
\newcommand{\lhat}{\hat{\ld}}
\newcommand{\qhat}{\hat{q}}

\renewcommand{\vec}[1]{[#1]}
\newcommand{\x}{\mathbf{x}}
\newcommand{\y}{\mathbf{y}}
\newcommand{\w}{\mathbf{w}}
\newcommand{\FM}{\textbf{J}}
\newcommand{\FMinv}{\textbf{J}^{-\!1}}

\newcommand{\trans}{^T}
\newcommand{\expec}[1]{\mathsf{E}\!\left\{#1\right\}}
\newcommand{\define}{\triangleq}

\renewcommand{\P}{P(\g)}
\newcommand{\Pn}{P_n(\ld)}
\newcommand{\I}{I(\g)}
\newcommand{\In}{I_n(\ld)}
\newcommand{\pY}[1]{p_Y(#1)}
\newcommand{\lnpY}{\ln p_Y(\y)}
\newcommand{\LLRn}{\Lambda_n(\ld)}
\newcommand{\Cosh}[1]{\cosh\!\left(#1\right)}
\newcommand{\Tanh}[1]{\tanh\!\left(#1\right)}
\newcommand{\dfdx}[2]{\frac{\partial #1}{\partial #2}}
\newcommand{\dfdfdxdy}[3]{\frac{\partial^2 #1}{\partial #2\partial #3}}
\newcommand{\dfdfdxdx}[2]{\frac{\partial^2 #1}{\partial #2^2}}
\newcommand{\fg}{f(\g)}
\newcommand{\hg}[1]{h(#1)}
\newcommand{\hginv}[1]{h^{-\!1}\!\left(#1\right)}
\newcommand{\gmusg}{g(\mu,\sg)}
\newcommand{\Q}[1]{Q\!\left({#1}\right)}
\newcommand{\J}[1]{J\!\left({#1}\right)}
\newcommand{\Exp}[1]{e^{#1}}
\newcommand{\Ln}[1]{\ln\!\left({#1}\right)}
\newcommand{\Logtwo}[1]{\log_2\!\left({#1}\right)}

\newcommand{\CRLB}{\Gamma}
\newcommand{\CRLBd}{\Gamma_\delta}
\newcommand{\CRLBmu}{\Gamma_\mu(\g)}
\newcommand{\CRLBsg}{\Gamma_\sg(\g)}
\newcommand{\CRLBld}{\Gamma_\ld(\g)}
\newcommand{\CRLBg}{\Gamma_\g(\g)}
\newcommand{\CRLBP}{\Gamma_P(\g)}

\newcommand{\Int}[5]{\hspace{-#1mm}\int\limits_{#2}^{#3}\hspace{-#1mm}#4\,d#5}
\newcommand{\MI}[2]{I(#1;#2)}
\newcommand{\Es}{E_s}
\newcommand{\Eb}{E_b}

\begin{document}

\title{Non-Data-Aided Parameter Estimation in an Additive White Gaussian Noise Channel}

\author{
\authorblockN{Fredrik Br\"{a}nnstr\"{o}m}
\authorblockA{Department of Signals and Systems\\
Chalmers University of Technology\\
SE-412 96 G\"oteborg, Sweden\\
Email: \texttt{fredrikb@s2.chalmers.se}\vspace{-10mm}}%
\and
\authorblockN{Lars K.~Rasmussen}
\authorblockA{Institute for Telecommunications Research\\
University of South Australia\\
Mawson Lakes, SA 5095, Australia\\
Email: \texttt{lars.rasmussen@unisa.edu.au}\vspace{-10mm}}%
}

\maketitle

\begin{abstract}
Non-data-aided (NDA) parameter estimation is considered for
binary-phase-shift-keying transmission in an additive white
Gaussian noise channel. Cram\'{e}r-Rao lower bounds (CRLBs) for
signal amplitude, noise variance, channel reliability constant and
bit-error rate are derived and it is shown how these parameters
relate to the signal-to-noise ratio (SNR). An alternative
derivation of the iterative maximum likelihood (ML) SNR estimator
is presented together with a novel, low complexity NDA SNR
estimator. The performance of the proposed estimator is compared
to previously suggested estimators and the CRLB. The results show
that the proposed estimator performs close to the iterative ML
estimator at significantly lower computational complexity.
\end{abstract}

\renewcommand{\thefootnote}{\empty}
\footnotetext{
    F.~Br\"{a}nnstr\"{o}m and L.~K.~Rasmussen are supported by the
    Swedish Research Council under Grant 621-2001-2976.
    F.~Br\"{a}nnstr\"{o}m is also supported by Personal Computing and
    Communication (PCC++) under Grant PCC-0301-09.
    L.~K.~Rasmussen is also supported by the Australian
    Government under ARC Grant DP0558861.
  }
\renewcommand{\thefootnote}{\arabic{footnote}}

\section{Introduction}

Most iterative decoders, e.g. turbo decoders
\cite{BerGlaThi93ICC}, rely on knowledge of the signal-to-noise
ratio (SNR) or the channel reliability constant
\cite{BerGlaThi93ICC,SumWil98TC,ReeAse97ICT}. The SNR is also
required for other functionalities in the receiver. Many SNR
estimators have been proposed, both data-aided (DA) -- that
require pilot symbols or feedback from the decoders
\cite{ReeAse97ICT,PauBea00TC,JesSam01VTC}, and non-data-aided
(NDA) -- that are only based on the received observables
\cite{SumWil98TC,PauBea00TC,WieGolMes02ICC}.

A comparison of both DA and NDA SNR estimators was performed in
\cite{PauBea00TC} and compared to the Cram\'{e}r-Rao lower bound
(CRLB) for DA estimators. The CRLB for NDA estimators was later
derived in \cite{Ala01CL}. The NDA maximum likelihood (ML)
estimator based on the expectation maximization (EM) algorithm was
proposed in \cite{WieGolMes02ICC} and also compared to other NDA
estimators. This NDA ML estimator was found iteratively, but
unfortunately requiring processing of all observables for each
iteration, making it computationally complex.

The contributions in this paper are as follows. To complement the
NDA CRLB for SNR in \cite{Ala01CL}, we derive the NDA CRLB for the
signal amplitude, the noise variance, the channel reliability
constant, and the bit-error rate (BER). It is also shown how to
estimate the \textit{a priori} probability of the transmitted
symbols, in the case when they are not equally likely.
Furthermore, we provide a more direct, alternative derivation of
the  NDA ML estimator and we propose a new, low complexity NDA SNR
estimator. The performance of the new estimator is compared to
previously suggested NDA estimators and found to be similar to
that of the NDA ML estimator. This performance is achieved with
significantly lower computational complexity than the ML
estimator. Only binary-phase-shift-keying (BPSK) transmission is
considered here, but generalization to $M$-PSK is straightforward
\cite{PauBea00TC,Ala01CL}.

\section{Problem Statement}

Let $X\in\{-1,+1\}$ denote a binary random variable with equally
likely symbols. Further, let $W$ represent a zero-mean Gaussian
random variable with unit variance. Define a new random variable $Y$
according to
\begin{align}
  Y\define\mu X + \sg W,
  \label{eq_Y}
\end{align}
with probability density function (PDF) expressed as
\cite{Ala01CL,WieGolMes02ICC}
\begin{align}
  \pY{y}&=\frac{1}{\sqrt{2\pi \sg^2}}\frac{1}{2}\left(\Exp{-\frac{(y-\mu)^2}{2\sg^2}}+\Exp{-\frac{(y+\mu)^2}{2\sg^2}}\right)\nonumber\\%
  &=\frac{1}{\sqrt{2\pi \sg^2}}\Exp{-\frac{\mu^2}{2\sg^2}}\Exp{-\frac{y^2}{2\sg^2}}\Cosh{\frac{\mu y}{\sg^2}}.%
  \label{eq_pY}
\end{align}

Let $x$, $y$, and $w$ denote samples from $X$, $Y$, and $W$,
respectively. $N$ independent samples of $Y$ is observed and
collected in a column vector $\y=\vec{y_1,y_2,\dots,y_N}\trans$.
If $\mu=\sqrt{\Es}$ and $\sg^2=N_0/2$, the model in \eqref{eq_Y}
represents BPSK transmission in additive white Gaussian noise
(AWGN)
\begin{align}
  \y=\mu \x + \sg \w,
  \label{eq_y}
\end{align}
where $\y$ is the matched filter output,
$\x=\vec{x_1,x_2,\dots,x_N}\trans$ the transmitted data, and
$\w=\vec{w_1,w_2,\dots,w_N}\trans$ white Gaussian noise. $\Es$ is
the transmitted energy and $N_0/2$ is the double-sided noise power
spectral density. Define the SNR as
\begin{align}
  \g\define \frac{\Es}{N_0}=\frac{\mu^2}{2\sg^2}.
  \label{eq_g}
\end{align}
Since all samples in $\y$ are assumed independent, the logarithm of
their joint PDF is given by \cite{Ala01CL}
\begin{align}
  \lnpY&=\Ln{\prod_{n=1}^N \pY{y_n}}
  \label{eq_lnpY}\\
  &\hspace{-10mm}=-\frac{N}{2}\Ln{2\pi\sg^2}-\!\frac{N\mu^2}{2\sg^2}-\!\!\sum_{n=1}^N \frac{y_n^2}{2\sg^2}+\!\!\sum_{n=1}^N \Ln{\Cosh{\!\frac{\mu y_n}{\sg^2}\!}\!}\!.%
  \nonumber
\end{align}

The average BER can be expressed as
\begin{align}
  \P\define \Pr(Y<0|X=+1)=\Q{\sqrt{2\g}},
  \label{eq_P}
\end{align}
where $Q(\alpha)\define
1/\sqrt{2\pi}\int_{\alpha}^{\infty}{\Exp{-\beta^2/2}}{d \beta}$ is
the Gaussian $Q$-function. The average mutual information (MI)
\cite{ten01TC} between $X$ and $Y$ in \eqref{eq_Y} can be
expressed as
\begin{align}
  \I\define \MI{X}{Y}=\J{\sqrt{8\g}},
  \label{eq_IXY}
\end{align}
where $J$ is defined as \cite{ten01TC}
\begin{align}
  \J{\alpha}\define 1-\frac{1}{\sqrt{2 \pi \alpha^2}}\Int{0}{-\infty}{+\infty}{\Logtwo{1+\Exp{-\beta}}\Exp{-\frac{(\beta-\alpha^2/2)^2}{2 \alpha^2}}}{\beta}.%
  \label{eq_J}
\end{align}
The log-likelihood ratio (LLR) for $y_n$ is defined as
\cite{ten01TC}
\begin{align}
  \LLRn\define \Ln{\frac{\pY{y_n|x_n=+1}}{\pY{y_n|x_n=-1}}}=\ld\, y_n, %
  \label{eq_LLR}
\end{align}
where $\ld$ is the channel reliability constant
\cite{BerGlaThi93ICC,ReeAse97ICT}
\begin{align}
  \ld\define\frac{2\mu}{\sg^2}.
  \label{eq_ld}
\end{align}
The instantaneous BER for a specific received symbol at position
$n$ can be estimated by \cite{HoeLanSor00ISTC}
\begin{align}
  \Pn\define\frac{1}{1+\Exp{|\ld y_n|}}.
  \label{eq_Pn}
\end{align}
The corresponding instantaneous MI for a symbol at position $n$ can
be estimated by \cite{Bra04PHD}
\begin{align}
  \In\define 1-2\frac{\Logtwo{1+\Exp{-\ld y_n}}}{1+\Exp{-\ld y_n}}.
  \label{eq_In}
\end{align}

With no knowledge of the transmitted symbols the average BER in
\eqref{eq_P} or the average MI in \eqref{eq_IXY} depend solely on
the SNR, $\g$. Also, in order to use an LLR-based decoder
\eqref{eq_LLR} (basically all turbo-like decoders
\cite{BerGlaThi93ICC,ten01TC} or soft decoders), to estimate the
instantaneous BER in \eqref{eq_Pn}, or to estimate the
instantaneous MI in \eqref{eq_In}, the channel reliability
constant in \eqref{eq_ld} needs to be known. However, as we show
in Section \ref{sec_Estimators}, the SNR and the channel
reliability constant are related through the second moment of the
observables. We therefore only need to estimate one of the two.
Here, we have chosen to estimate the SNR, $\gamma$.

\section{Cram\'{e}r-Rao Lower Bound}\label{sec_CRLB}

The CRLB, here denoted by $\CRLB$, provides a lower bound on the
variance of any unbiased estimator \cite{Kay93BK}. Let $\gmusg$
represent an arbitrary function of the parameters $\mu$ and $\sg$,
and define $\delta\define\gmusg$. The normalized CRLB (NCRLB) for
$\delta$ can then be calculated as \cite{Kay93BK}
\begin{align}
  \CRLBd\define\frac{1}{\delta^2}\left[%
  \begin{array}{cc}
    \!\!\!\dfdx{\gmusg}{\mu} & \!\dfdx{\gmusg}{\sg} \\
  \end{array}%
  \!\!\!\right]
  \FMinv
  \left[%
  \begin{array}{cc}
    \!\!\!\dfdx{\gmusg}{\mu} & \!\dfdx{\gmusg}{\sg} \\
  \end{array}%
  \!\!\!\right]
  \trans\!,
  \label{eq_CRLBd}
\end{align}
where $\FM$ is the Fisher information matrix \cite{Kay93BK}, defined
as
\begin{align}
  \FM\define
  \left[%
  \begin{array}{cc}
  -\expec{\dfdfdxdx{\lnpY}{\mu}     } & -\expec{\dfdfdxdy{\lnpY}{\mu}{\sg}} \\
  -\expec{\dfdfdxdy{\lnpY}{\sg}{\mu}} & -\expec{\dfdfdxdx{\lnpY}{\sg}} \\
  \end{array}%
  \right].
  \label{eq_FM}
\end{align}
Here $\expec{\cdot}$ denotes the expectation over $Y$. A similar
Fisher information matrix as in \eqref{eq_FM} has been derived in
\cite{Ala01CL} and the inverse can be expressed as
\begin{align}
  \FMinv=\frac{\sg^2}{N}\frac{1}{2-2\fg-8\g\fg}
  \left[%
  \begin{array}{cc}
  2-8\g\fg & \sqrt{8\g}\fg \\
  \sqrt{8\g}\fg &   1-\fg \\
  \end{array}%
  \right],
  \label{eq_FMinv}
\end{align}
where $\fg$ is a function of $\g$ \cite{Ala01CL}
\begin{align}
  \fg\define \frac{\Exp{-\g}}{\sqrt{2\pi}}\Int{0}{-\infty}{+\infty}{\frac{\beta^2 \Exp{-\frac{\beta^2}{2}}}{\Cosh{\beta\sqrt{2\g}}}}{\beta}.%
  \label{eq_fg}
\end{align}

Using \eqref{eq_CRLBd}, the CRLB for $\g$ can be calculated as
reported in \cite{WieGolMes02ICC} and \cite{Ala01CL}. The NCRLBs for
$\mu$, $\sg$, $\g$, $\ld$, and $\P$ are
\begin{align}
  \CRLBmu&=\frac{1}{\g N}\frac{1-4\g\fg}{2-2\fg-8\g\fg}\label{eq_CRLBmu},\\%
  \CRLBsg&=\frac{1}{N  }\frac{1-\fg}{2-2\fg-8\g\fg}\label{eq_CRLBsg},\\%
  \CRLBg &=\frac{1}{\g N}\frac{4+4\g-4\g\fg}{2-2\fg-8\g\fg}\label{eq_CRLBg},\\%
  \CRLBld&=\frac{1}{\g  N}\frac{1+4\g}{2-2\fg-8\g\fg}\label{eq_CRLBld},\\%
  \CRLBP &=\frac{1}{\P^2}\left(\dfdx{\P}{\g}\right)^2 \g^2 \CRLBg\label{eq_CRLBP}\\
  &=\frac{\Exp{-2\g}}{\pi N \Q{\sqrt{2\g}}^2}\frac{1+\g-\g\fg}{2-2\fg-8\g\fg}.\nonumber%
\end{align}
The NCRLB for $\I$ can also be found in a similar way by replacing
$\P$ in \eqref{eq_CRLBP} with $\J{\sqrt{8\g}}$. Unfortunately,
there is no simple form to express $\dfdx{\J{\sqrt{8\g}}}{\g}$.
Note that $\FMinv$ for DA estimation is found by setting $\fg=0$
in \eqref{eq_FMinv} \cite{Ala01CL}. This implies that the NCRLBs
for DA estimation of $\mu$, $\sg$, $\g$, $\ld$, and $\P$ are
easily found by letting $\fg=0$ in \eqsref{eq_CRLBmu}{eq_CRLBP}.

\section{Estimators}\label{sec_Estimators}

Define moment $k$ of $Y$ as
\begin{align}
  M_k&\define \expec{Y^k}\approx \frac{1}{N}\sum_{n=1}^N y_n^k,
  \label{eq_Mk}
\end{align}
which can be approximated by its sample average \cite{PauBea00TC}.
The second moment of $Y$ is
\begin{align}
  M_2=\mu^2+\sg^2.
  \label{eq_M2}
\end{align}
Assume that an estimate of $\g$ exists, denoted by $\ghat$.
Combining \eqref{eq_M2} with \eqref{eq_g} and \eqref{eq_ld} gives
the following estimators for $\mu$, $\sg$, and $\ld$
\begin{align}
  \mhat&=\sqrt{\frac{2\ghat M_2}{1+2\ghat}},
  \label{eq_mhat}\\
  \shat&=\sqrt{\frac{M_2}{1+2\ghat}},
  \label{eq_shat}\\
  \lhat&=\sqrt{\frac{8\ghat+16\ghat^2}{M_2}}.
  \label{eq_lhat}
\end{align}
The next sub-sections present different estimators for $\g$ that
can be used to estimate the above parameters.

\subsection{Conventional Method}

The absolute moment (AM) of $Y$ is defined as
\cite{Pie98JSC,SumWil98TC}
\begin{align}
  A\define\expec{|Y|}=\mu+\sg\sqrt{\frac{2}{\pi}}\Exp{-\frac{\mu^2}{2\sg^2}}-2\mu \Q{\frac{\mu}{\sg}},%
  \label{eq_AM}
\end{align}
and can also be approximated by the sample average \cite{PauBea00TC}
\begin{align}
  A\approx \frac{1}{N}\sum_{n=1}^N |y_n|.
  \label{eq_A}
\end{align}
For large $\mu$ or small $\sg$, the AM will tend to $\mu$
\begin{align}
  \lim_{\sg\rightarrow 0} A=  \lim_{\mu\rightarrow \infty} A= \mu.
  \label{eq_limA}
\end{align}
In other words, for high values of $\g$, $\ghat$ can be closely
approximated by
\begin{align}
  \ghat=\frac{A^2}{2(M_2-A^2)}=\frac{\frac{A^2}{M_2}}{2\left(1-\frac{A^2}{M_2}\right)},
  \label{eq_ghat_CM}
\end{align}
using \eqref{eq_g} and \eqref{eq_M2}. This estimator was first
introduced in \cite{Gil66JPL} and will here be referred to as the
conventional method (CM) estimate.

\subsection{Maximum Likelihood}

The ML estimator maximizes the joint PDF in \eqref{eq_lnpY}. Taking
the partial derivatives of $\lnpY$ gives
\begin{align}
  \dfdx{\lnpY}{\mu}&=\frac{N}{\sg^2}\left(\frac{1}{N}\sum_{n=1}^N y_n \Tanh{\frac{\mu
  y_n}{\sg^2}}-\mu\right),
  \label{eq_dlnpdmu}\\
  \dfdx{\lnpY}{\sg}&=
  \label{eq_dlnpdsg}\\
  &\hspace{-6mm}\frac{N}{\sg^3}\!\left(\mu^2-\sg^2+\frac{1}{N}\sum_{n=1}^N y_n^2-\frac{2\mu}{N}\sum_{n=1}^N y_n \Tanh{\frac{\mu y_n}{\sg^2}}\!\!\right)\!.%
  \nonumber
\end{align}
Setting the derivatives in \eqref{eq_dlnpdmu} and \eqref{eq_dlnpdsg}
to zero and solve for $\sg^2$ gives
\begin{align}
  \sg^2=\frac{1}{N}\sum_{n=1}^N y_n^2 - \mu^2=M_2-\mu^2.
  \label{eq_sg2}
\end{align}
Inserting \eqref{eq_sg2} in \eqref{eq_dlnpdmu} gives an expression
depending only on $\y$, $\mu$ and $M_2$, which can be solved
iteratively by
\begin{align}
  \mhat_{k+1}=\frac{1}{N}\sum_{n=1}^N y_n \Tanh{\frac{\mhat_k\, y_n}{M_2-\mhat_k^2}},%
  \label{eq_mu_it}
\end{align}
where $\mhat_{k}$ denotes the estimate of $\mu$ after $k$
iterations. The iteration in \eqref{eq_mu_it} is identical to the
iteration in the EM algorithm presented in \cite{WieGolMes02ICC},
but here derived in a different way. A good starting point for the
iterative estimator is the CM estimate of $\mu$, $\mhat_0=A$.
After $K$ iterations the SNR can be estimated by
\begin{align}
  \ghat=\frac{\mhat_K^2}{2(M_2-\mhat_K^2)},
  \label{eq_ghat_ML}
\end{align}
which will be referred to as the ML estimator.

\subsection{Method of Moments}

The approach of estimating a parameter based on the moments of the
observables is known as the method of moments (MM) \cite{Kay93BK}.
The fourth moment of $Y$ is \cite{PauBea00TC}
\begin{align}
  M_4&\define \expec{Y^4} = \mu^4 + 6\mu^2\sg^2 + 3 \sg^4.
  \label{eq_M4}
\end{align}
Combining \eqref{eq_M4} with \eqref{eq_M2} gives the MM estimator,
e.g. \cite{PauBea00TC},
\begin{align}
  \ghat=\frac{\sqrt{6 M_2^2-2 M_4}}{4 M_2-2\sqrt{6 M_2^2-2 M_4}},
  \label{eq_ghat_MM}
\end{align}
where $M_2$ and $M_4$ are approximated by the sample average
\eqref{eq_Mk}. If $M_4>3M_2^2$, \eqref{eq_ghat_MM} is no longer
real and $\ghat$ is set to zero. This will be referred to as the
MM estimator.

\subsection{Absolute Moment}

An estimator based on the second moment and the AM can be found by
combining \eqref{eq_M2} with \eqref{eq_AM}. Unfortunately, there
is no closed-form analytical solution for $\mu$ and $\sg$ as for
the MM estimator. However, dividing the square of $A$ by $M_2$
gives an expression that only depends on $\g$,
\begin{align}
  \hg{\g}\define\frac{A^2}{M_2}=\frac{2\g}{2\g+1}\!\left(\!1+\frac{1}{\sqrt{\pi\g}}\Exp{-\g}-2 \Q{\!\sqrt{2\g}}\!\!\right)^2\!\!.%
  \label{eq_hg}
\end{align}
An estimator for $\g$ can therefore be stated as
\begin{align}
  \ghat=\hginv{\frac{A^2}{M_2}}.
  \label{eq_ghat_AM}
\end{align}
Since there is no closed-form solution to \eqref{eq_ghat_AM},
alternative methods must be explored. In \cite{Pie98JSC}, a
table-lookup for $\hginv{\alpha}$ is suggested. A different
approach is to approximate $\hginv{\alpha}$ with a simple
closed-form function, which was done in \cite{SumWil98TC} as,
\begin{align}
  \hginv{\alpha}\approx \frac{1}{2}10^{(-34.0516/\alpha^2+65.9548/\alpha-23.6184)/10}.%
  \label{eq_hginv_P2}
\end{align}
The estimator in \eqref{eq_ghat_AM}, using the approximation in
\eqref{eq_hginv_P2} is referred to as the second-order polynomial
(P2) estimator.

From \eqref{eq_hg} it is easy to verify that
\begin{align}
  \hg{0}=\frac{2}{\pi}\approx 0.6366, \quad\textrm{ and }\quad \hg{\infty}=1.
  \label{eq_hg_0_infty}
\end{align}
Therefore, we suggest the following approximation of $\hg{\alpha}$
and its inverse
\begin{align}
  \hg{\alpha}&\approx
  1-\left(1-\frac{2}{\pi}\right)\!\left(H_1\alpha^{H_2}+1\right)^{H_3},
  \label{eq_hg_approx}\\
  \hginv{\alpha}&\approx\left(\frac{\left(\frac{1-\alpha}{1-\frac{2}{\pi}}\right)^{1/H_3}-1}{H_1}\right)^{1/H_2}.
  \label{eq_hginv_approx}
\end{align}
Numerical optimization, using the Nelder-Mead simplex method
\cite{NelMea65CJ} to minimize the mean squared difference between
\eqref{eq_hg} and \eqref{eq_hg_approx} gives $H_1=0.6153$,
$H_2=1.5296$, and $H_3=-0.6575$. The estimator in
\eqref{eq_ghat_AM}, using the novel approximation in
\eqref{eq_hginv_approx}, with $\ghat=0$ whenever $A^2/M_2\le 2/\pi$,
is a new approach we propose and is here referred to as the AM
estimator.

\begin{figure}[t]
  \setlength{\unitlength}{0.85mm}
  \centering
  \small
  \begin{picture}(94,84)(-1,-2)
    \put(-3,0){ \includegraphics[width=85mm]{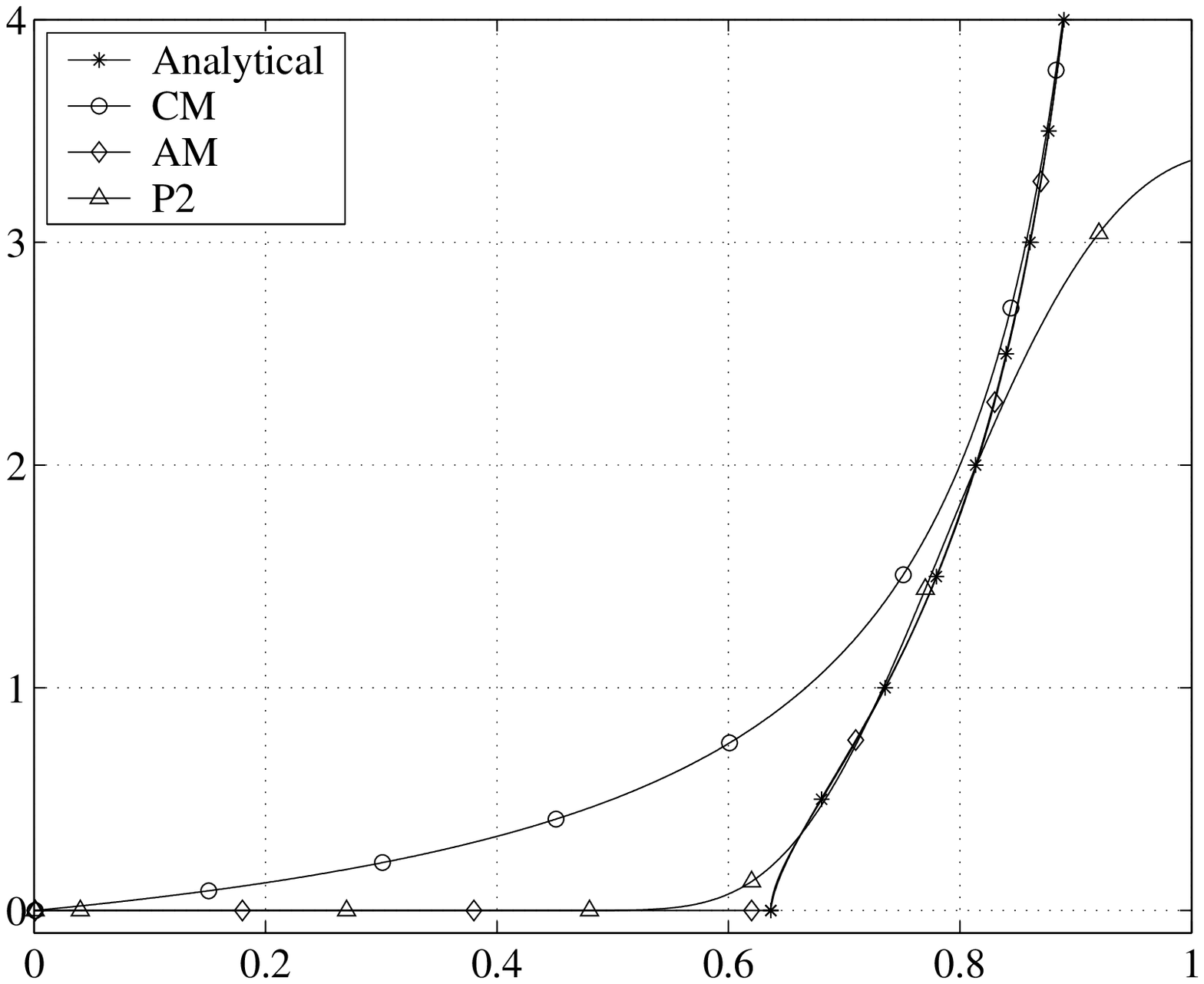}}
    \put(-2,46){\makebox(0,0)[r]{\rotatebox{90}{$\ghat=\hginv{A^2/M_2}$}}}
    \put(49,-1){\makebox(0,0)[t]{$A^2/M_2$}}
  \end{picture}
  \caption{Different approximations of $\ghat=\hginv{A^2/M_2}$.}
  \label{fig_h_approx}
\end{figure}

\figref{fig_h_approx} shows the analytical expression from
\eqref{eq_hg} together with its indistinguishable approximation in
\eqref{eq_hg_approx}. Since the CM estimator in \eqref{eq_ghat_CM}
depends only on $A^2/M_2$ it is also shown in
\figref{fig_h_approx}. It is clear that the function used by the
CM estimator converges to the analytical one for large $\g$, but
differs for small $\g$. The same figure also shows the
approximation in \eqref{eq_hginv_P2}. Since this polynomial
approximation was only optimized between $-3$ to $3$ dB
($\g=0.5$--$2$) \cite{SumWil98TC} it differs from the analytical
expression outside this region.

\subsection{Non-Equiprobable Symbols}

Define $q$ to be the \textit{a priori} probability of $X$ in
\eqref{eq_Y}
\begin{align}
  q\define\Pr(X=+1).
  \label{eq_q}
\end{align}
The ML estimator is invariant to non-equiprobable symbols and gives
the same results even if $q=1$ \cite{WieGolMes02ICC}. It is
straightforward to show that $A$, $M_2$, and $M_4$ are independent
of $q$ \cite{Bra04PHD}. Since the CM, the MM, the P2, and the AM
estimators are based only on these quantities, they will give the
same results independent of $q$.

However, when $q\ne 0.5$, the odd moments are non-zero,
\begin{align}
  M_1\define\expec{Y}=\mu(2 q -1)\approx \frac{1}{N}\sum_{n=1}^N y_n.
  \label{eq_M1}
\end{align}
This means that the \textit{a priori} probability $q$ can be
estimated using $M_1$, by combining \eqref{eq_M1} with \eqref{eq_g}
and \eqref{eq_M2}
\begin{align}
  \qhat&=\frac{M_1}{2}\sqrt{\frac{1+2\ghat}{2\ghat
  M_2}}+\frac{1}{2}.
  \label{eq_qhat}
\end{align}

\section{Numerical Examples}\label{sec_Examples}

The performance of the SNR estimators is evaluated based on their
normalized mean squared error (NMSE)
\begin{align}
  \frac{1}{L}\sum_{j=1}^L \frac{(\ghat_j-\g)^2}{\g^2}
  \label{eq_NMSE}
\end{align}
and their normalized bias (NB)
\begin{align}
  \frac{1}{L}\sum_{j=1}^L \frac{\ghat_j-\g}{\g},
  \label{eq_NB}
\end{align}
where $\ghat_j$ is estimated based on $N$ samples. The number of
trials was chosen to $L=100\,000$. The best estimator is an unbiased
estimator with minimum NMSE \cite{Kay93BK}.

\begin{figure}[t]
  \setlength{\unitlength}{0.88mm}
  \centering
  \small
  \begin{picture}(94,81)(1,-2)
    \put(-3,0){ \includegraphics[width=88mm]{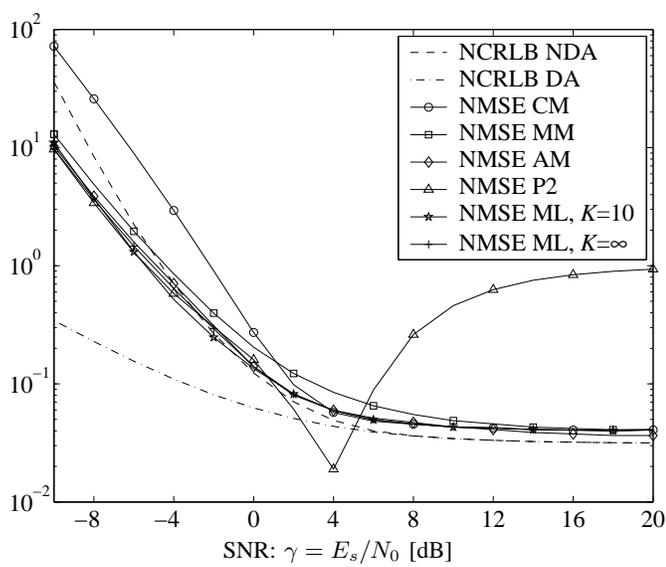}}
    \put(48,-1){\makebox(0,0)[t]{SNR: $\g=\Es/N_0$ [dB]}}
  \end{picture}
  \caption{Normalized MSE for $\ghat$ when $N=64$ samples are observed.}
  \label{fig_NMSE_g_gs}
  \vspace{-3mm}
\end{figure}

\begin{figure}[t]
  \setlength{\unitlength}{0.90mm}
  \centering
  \small
  \begin{picture}(94,81)(1,-2)
    \put(-3,0){ \includegraphics[width=88mm]{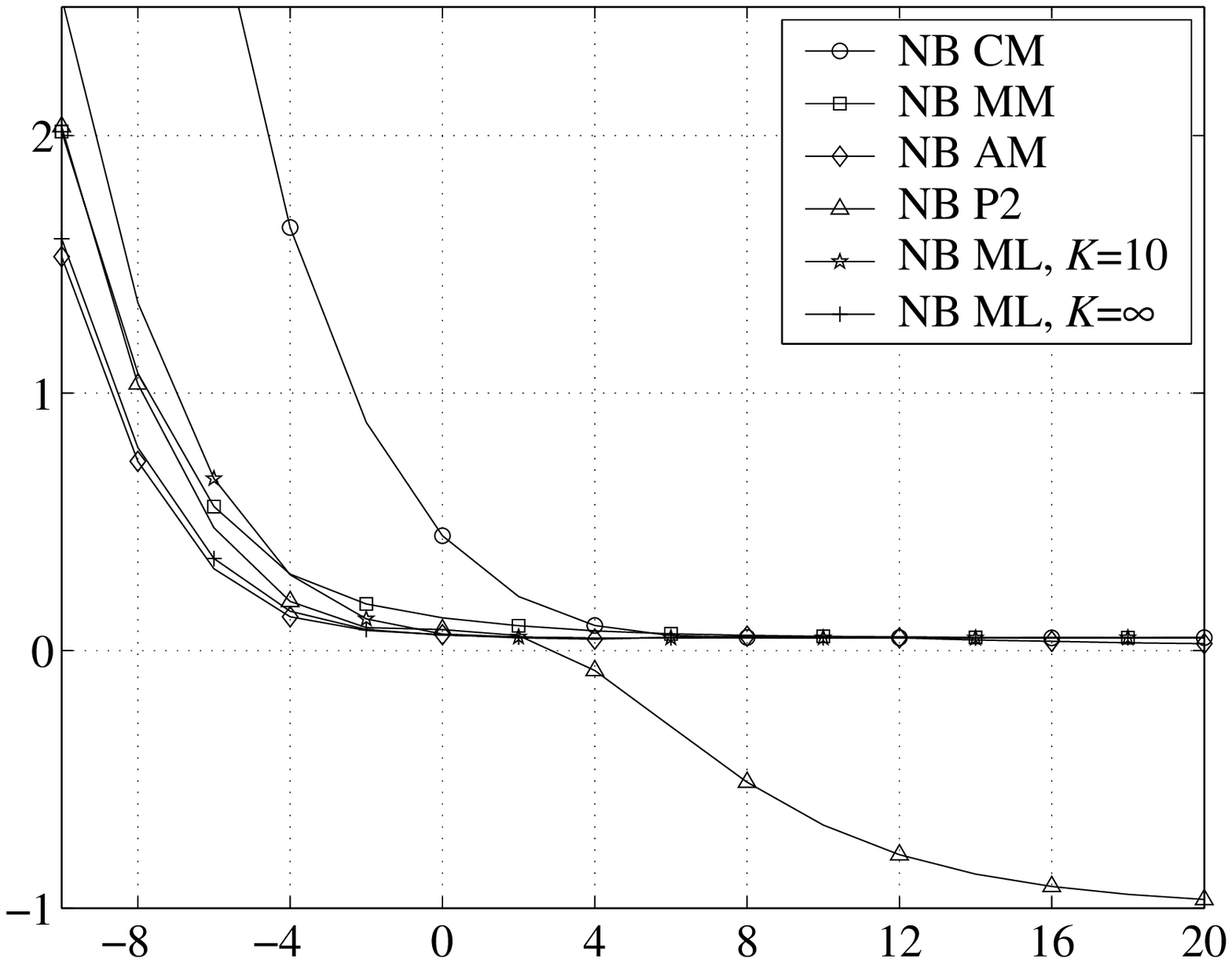}}
    \put(47,-1){\makebox(0,0)[t]{SNR: $\g=\Es/N_0$ [dB]}}
  \end{picture}
  \caption{Normalized bias for $\ghat$ when $N=64$ samples are observed.}
  \label{fig_NB_g_gs}
  \vspace{-2mm}
\end{figure}

\figref{fig_NMSE_g_gs} shows the NMSE and \figref{fig_NB_g_gs}
shows the NB, both for $N=64$ observables. The NDA and the DA
NCRLB are also included as a reference, even though they are only
bounds for unbiased estimators. All the estimators presented here
are biased when $N=64$, even for high $\g$ which is evident from
\figref{fig_NB_g_gs}. Different approaches to reduce the bias has
been suggested, e.g.,
\cite{PauBea00TC,JesSam01VTC,WieGolMes02ICC}. The CM estimator has
a large NB (and therefore also a large NMSE) for low $\g$. For
large $\g$ the CM estimator approaches the ML estimator, which was
shown analytically in \cite{LetGra03AusCTW}. In fact,
\figsref{fig_NMSE_g_gs}{fig_NB_g_gs} show that all estimators,
except the P2 estimator converges to the same constant NMSE and
constant NB for high $\g$ (the NB is around 5\% above the true
$\g$). The P2 estimator only works well between -3 to 3 dB, the
interval for which it was optimized. The MM estimator has the
second highest NMSE for low $\g$. The ML estimator after $K=10$
iterations has the lowest NMSE at -6 dB, but \figref{fig_NB_g_gs}
shows that it at the same time has the second highest NB. Finally,
the suggested AM estimator has almost identical performance (both
in NMSE and NB) as the ML estimator for all $\g$, even though it
has a computationally complexity that is less than the first ML
iteration.

\begin{figure}[t]
  \setlength{\unitlength}{0.90mm}
  \centering
  \small
  \begin{picture}(94,80)(1,-2)
    \put(-3,0){ \includegraphics[width=91mm]{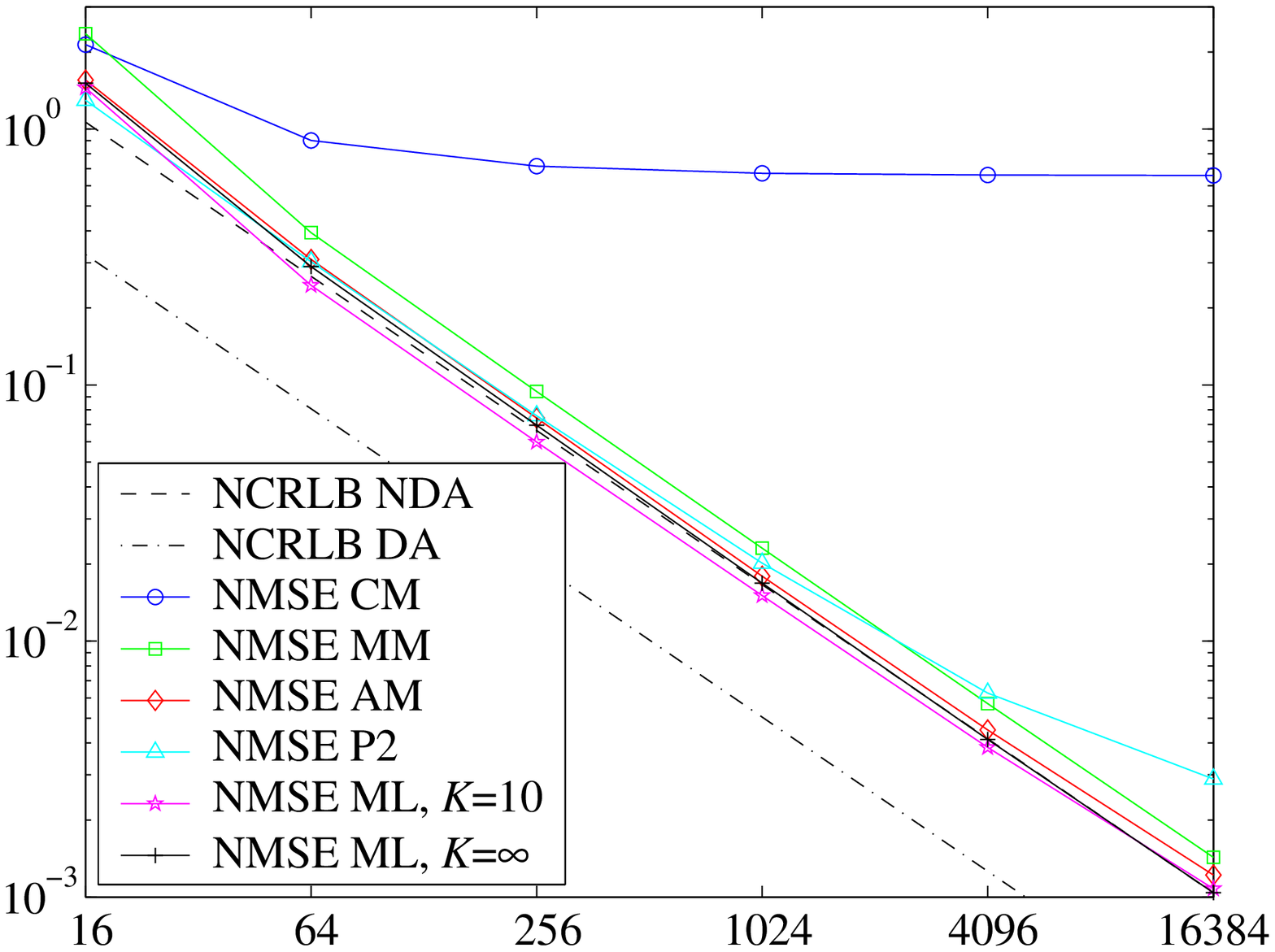}}
    \put(50,-1){\makebox(0,0)[t]{N}}
  \end{picture}
  \caption{Normalized MSE for $\ghat$ when $\g=-2$ dB.}
  \label{fig_NMSE_g_K}
  \vspace{-3mm}
\end{figure}

\figsref{fig_NMSE_g_K}{fig_NB_g_K} show the NMSE and the NB for
different $N$ at -2 dB. This corresponds to an $\Eb/N_0$ around 1
dB for a half-rate code, e.g. the original turbo code
\cite{BerGlaThi93ICC}. At this low SNR, the CM estimator has bad
performance. The NB saturates around 60\% above the true value
(not shown here), which gives the high NMSE in
\figref{fig_NMSE_g_K}. The P2 estimator has a negative NB for
large $N$ at this SNR. The ML estimator after $10$ iterations and
the AM estimator have a small positive NB (around 1\%) for large
$N$. The ML estimator and the MM estimator are the only two
estimators that are unbiased for large $N$, but only the ML
estimator approaches the NCRLB in \figref{fig_NMSE_g_K} which
makes it asymptotically optimal \cite{Kay93BK}. The second best
estimator, after the ML estimator, over the whole range of $N$ is
the suggested AM estimator.

\section{Conclusions}

In this paper we have derived the NDA NCRLB for the signal
amplitude, the noise variance, the channel reliability constant,
and the BER in an AWGN channel with BPSK modulated transmission.
It was also shown that these parameters, as well as the \textit{a
priori} probability of the transmitted symbols and the
instantaneous MI can all be estimated based on the SNR estimate. A
novel SNR estimator with low computationally complexity was
introduced and shown to be surpassed in performance only by the
iterative ML estimator among previously suggested estimators. The
proposed estimator performs close to the performance of the
iterative ML estimator at significantly lower computationally
complexity.

\begin{figure}[t]
  \setlength{\unitlength}{0.90mm}
  \centering
  \small
  \begin{picture}(94,81)(1,-2)
    \put(-3,0){ \includegraphics[width=91mm]{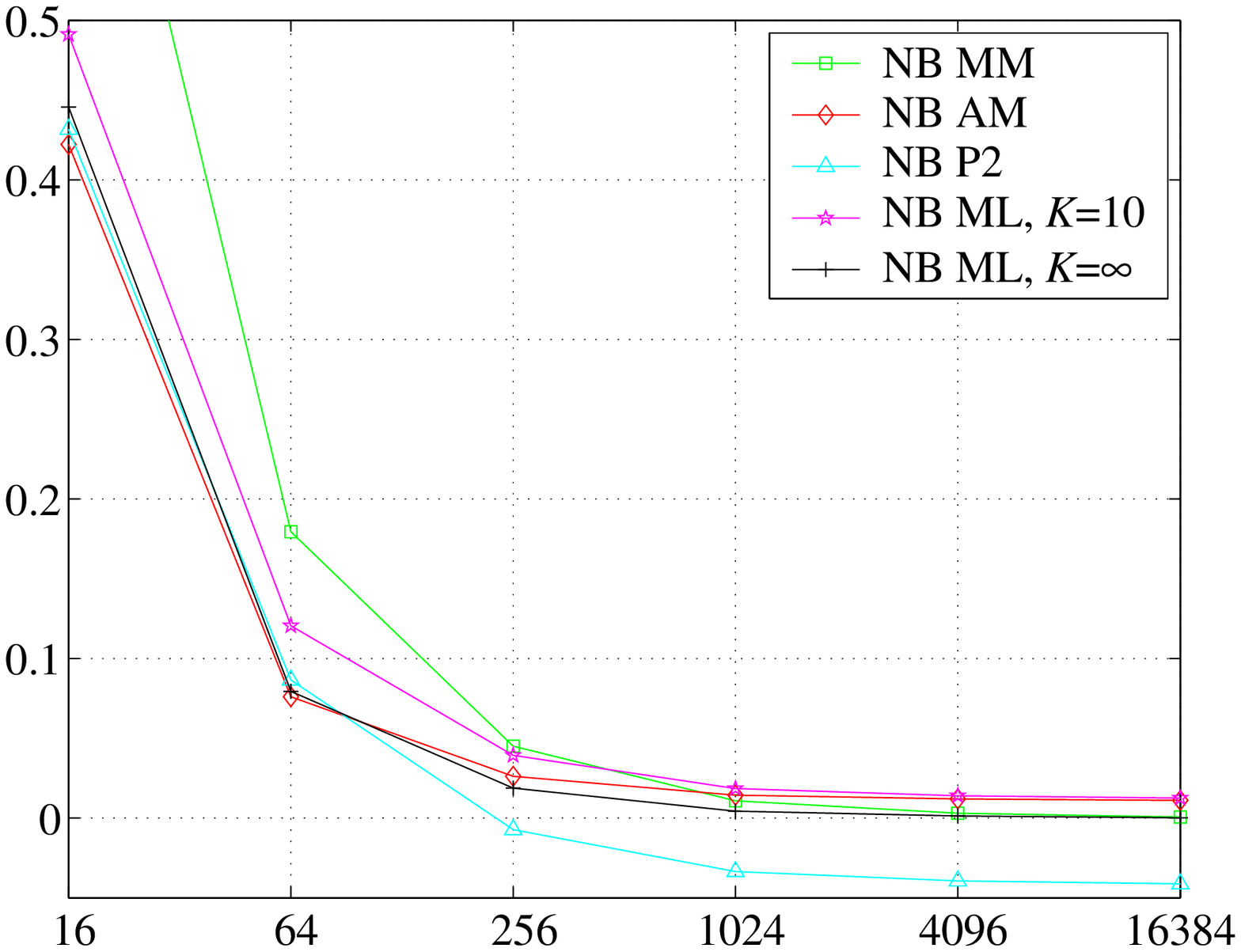}}
    \put(49,-1){\makebox(0,0)[t]{N}}
  \end{picture}
  \caption{Normalized bias for $\ghat$ when $\g=-2$ dB.}
  \label{fig_NB_g_K}
  \vspace{-2mm}
\end{figure}


\end{document}